\documentclass{iopjournal}

\usepackage{amsmath}
\usepackage{cite}
\usepackage{comment}

\begin{document}
	
    \articletype{Paper} 
	
    \title{Phase matching of high-order harmonics driven by a combined coherent and bright squeezed vacuum field} 

    \author{
    Yaoshun Sun$^{1,2}$,
    Shang Ma$^{1,2}$,
    Xintong He$^{1,2}$, 
    Jiaao Cai$^{1,2}$, 
    Wentao Wang$^{1,2}$, 
    Mingxuan Li$^{1,2}$\orcid{0000-0003-1186-3314},\\
    Xiangyu Tang$^{1,2*}$\orcid{0009-0008-1597-1842}, 
    Sizuo Luo$^{1,2*}$\orcid{0000-0002-8969-5958}
    }
    
    \affil{
    $^1$Institute of Atomic and Molecular Physics, Jilin University, 130012, Changchun, China.\\
	$^2$Jilin Provincial Key Laboratory of Transient Quantum Process Control and Application, Jilin University, Changchun 130012, China.
    }
    

    \email{tangxiangyu@jlu.edu.cn; luosz@jlu.edu.cn}
	
	
	\begin{abstract}
    High-order harmonic generation (HHG) driven by non-classical light provides new insights into light-matter interactions and paves the way to quantum attosecond spectroscopy. Achieving phase matching is vital for generating bright HHG sources, while it has received comparatively limited attention in quantum-light-driven HHG. In this work, we investigate the phase matching of HHG driven by an intense coherent field combined with a weak bright squeezed vacuum (BSV) field, using a semiclassical ensemble of coherent-field realizations sampled from the Husimi-$Q$ probability distribution. We reveal the physical mechanism behind the distinct pressure dependence of the harmonic yields for odd, even and half-integer sidebands spectral components, as observed in our previous experiment [\href{https://doi.org/10.48550/arXiv.2607.20941}{	arXiv:2607.20941}]. The simulations and phase-matching model demonstrate that the BSV-induced perturbation of the electron action phase could introduce an additional sub-cycle  phase mismatch, such that the amplitude of sub-cycle bursts could be significantly modified by gas pressure after macroscopic propagation, even when the harmonic yield is near the optimum. The resulting amplitude asymmetry of sub-cycle bursts leads to a significant intensity contrast between half-integer sidebands harmonics, manifesting as distinct pressure dependence of their yields. Our findings reveal the phase-matching mechanism in two-color quantum-light-driven HHG, and suggest opportunities for macroscopic control of the non-classical HHG source.
	\end{abstract}

    \section{Introduction}

    High-order harmonic generation (HHG) resulting from the interaction between intense femtosecond laser pulses with matter provides coherent radiation that spans the extreme-ultraviolet and soft-X-ray regions\cite{McPherson1987,Ferray1988,L’Huillier1993,Popmintchev2012}. In gases, its microscopic mechanism is understood within the three-step model: an electron is released through tunneling ionization, subsequently accelerated by the laser field, and finally driven back to its parent ion, where recombination leads to the emission of a high-energy photon\cite{Corkum1993,Lewenstein1994}. The resulting phase-locked harmonics have underpinned attosecond pulses generation and its spectroscopy\cite{Dinu2003,Kienberger2004,Hentschel2001,Krausz2009,Isinger2017,Gruson2016,Li2025a,Li2025,Jiang2024}. Conventional HHG universally treats the driving pulse as a deterministic classical field. Although this approximation is appropriate for coherent light, it excludes photon-number and field-quadrature fluctuations as dynamical variables. The quantum states of non-classical light can exhibit photon statistics and field fluctuations that differ markedly from those of coherent light. Therefore, quantum driving light is emerging as a new frontier in strong-field physics, attosecond science and quantum optics\cite{Lewenstein2021,Stammer2023,EvenTzur2023,RiveraDean2025,RiveraDean2026,Laurell2025,Tsatrafyllis2019,Sennary2025,Bhattacharya2023}. In strong-field interactions, theory has predicted that the statistics of the driving light can modify ionization, electron dynamics, and harmonic emission\cite{Gorlach2020,Stammer2022,Tzur2024,Liu2025,Lyu2025,Li2026,Liu2026,Tzur2025,Long2025,RiveraDean2025a,RiveraDean2025b,Wang2024,EvenTzur2024,Gorlach2023,Fang2023}.
    The bright squeezed vacuum (BSV) field, typically generated by parametric down-conversion (PDC) or four-wave-mixing process\cite{Lyu2026,Iskhakov2012,Sharapova2015,Liu2016,Vaidya2020}, is especially attractive because it can link the quantum optics and strong-field physics due to its large intensity fluctuation. Recent experiments have used BSV field to investigate multiphoton electron emission, high-harmonic generation in solids, and atomic tunneling ionization\cite{Heimerl2024,Rasputnyi2024,Lemieux2025,Jiang2026,Heimerl2025}.
    
    However, the limited low pulse energy due to low conversion efficiency of PDC constrains the direct gas-phase HHG using a BSV field. A practical alternative is a hybrid two-color driving field scheme, in which a strong coherent pulse drives the principal ionization--recollision dynamics while a weak BSV field opens fluctuation-sensitive photon-exchange pathways\cite{Tzur2025,Tzur2025a}. The hybrid quantum driving field has also enabled control of photon bunching in harmonic emission. For a coherent field combined with a BSV field at half the frequency, the single-photon process produces half-integer sidebands harmonics, whereas the two-photon process results in even-order components and can retain signatures of the driving-field quantum fluctuations.
	These developments motivate an understanding of how fluctuation-dependent microscopic harmonic emission builds up during macroscopic propagation. In experiments, the measured HHG spectrum is formed by the coherent buildup of radiation throughout the medium, and governed by propagation effects including diffraction, absorption, neutral-atom, and plasma dispersion\cite{L’Huillier1990,Balcou1993,Dachraoui2009,Gaarde2008,Jin2011}. Therefore, the microscopic spectra alone cannot predict the experimentally observed signals completely. Macroscopic propagation or phase-matching condition can act as a filter by suppressing a strong single-atom channel through destructive interference or enhancing a weak channel through coherent accumulation\cite{Merdji2006,He2015}. Thus, a favorable phase-matching condition is vital for the access and manipulation of bright HHG sources\cite{Popmintchev2009,Chen2010,Constant1999}. 

    In this work, we investigate macroscopic HHG driven by a two-color hybrid quantum driving field consisting of a strong 800-nm coherent field and a weak 1600-nm BSV field. We show the spectral components of odd, even, and satellite sidebands harmonics could exhibit distinct dependence of harmonic yield on gas pressure. The role of macroscopic propagation in the modulation of spectral structure is revealed, which differs from that at the microscopic level. Phase-matching model using a semiclassical description shows that the quantum fluctuation of BSV field would introduce shot-dependent, sub-cycle-resolved phase mismatch. It consequently results in the distinct buildup dynamics for the sub-cycle bursts as well as spectral components. The significant amplitude asymmetry of sub-cycle burst after macroscopic propagation is responsible for the modulation of spectral structure, manifesting as the different dependence of harmonic yield on gas pressure among the components. Our findings establish a framework for the phase-matching analysis of HHG driven by a two-color quantum light field, and demonstrate the possibility of macroscopic control of non-classical HHG.

	\section{Methods}
	\subsection{Simulation of two-color quantum driving field}\label{sec:2_1}

	Within the quasi-monochromatic approximation, valid for narrow-bandwidth laser pulses, the coherent driving field can be written as
	\begin{equation}\label{Eq:coh_Est_revised}
		E_{\mathrm{coh}}(\mathbf{s},t)
		=
		\left|\widetilde{u}(\mathbf{s})\right|
		\cos\!\left[
		\omega t+\varphi(\mathbf{s})
		\right]f(t),
	\end{equation}
	where
	$\widetilde{u}(\mathbf{s})
	=
	\left|\widetilde{u}(\mathbf{s})\right|
	e^{i\varphi(\mathbf{s})}$
	is the complex spatial electric-field,
	and $f(t)$ is the temporal envelope.
	For a Gaussian beam propagating along the $z$ direction, the spatial mode in cylindrical coordinates is written as
	\begin{equation}\label{Eq:Gaussian_mode_revised}
		\widetilde{u}(r,z)
		=
		E_0\frac{Q(0)}{Q(z)}
		\exp\!\left[
		-\frac{ikr^2}{2Q(z)}
		\right],
	\end{equation}
	where $E_0$ is the peak electric-field amplitude. $Q(z)=z+iz_{\mathrm{R}}$,
	and $z_{\mathrm{R}}=kW^2/2$	is the Rayleigh length.	Here, $W$ denotes the beam waist and $k=\omega/c$ is the wave numbers.
	The field in Eq.~\eqref{Eq:Gaussian_mode_revised} is expressed in the moving frame $(z=z',t=t'-z'/c)$ where the phase factor $e^{-ikz}$ is omitted.
	
    We consider a BSV field occupying a single mode and adopt a semiclassical approach based on its Husimi-$Q$ distribution described by 
    \begin{equation}\label{Eq:BSV_Phase_Space}
		\mathcal{P}_{\mathrm{BSV}}(x,y)
		=
		\frac{1}{2\pi\cosh r}
		\exp\!\left[
		-\frac{x^2}{1+e^{-2r}}
		-\frac{y^2}{1+e^{2r}}
		\right]
		=
		\frac{1}
		{2\pi\sigma_{\mathrm{x}}\sigma_{\mathrm{y}}}
		\exp\!\left[
		-\frac{x^2}{2\sigma_{\mathrm{x}}^2}
		-\frac{y^2}{2\sigma_{\mathrm{y}}^2}
		\right].
    \end{equation}    
    For a squeezing parameter $r$, each field realization is specified by two independent, zero-mean Gaussian variables $x$ and $y$ with variances
    \begin{equation}
    \left\langle x^2\right\rangle
    =\sigma_x^2=\frac{1+e^{-2r}}{2},
    \qquad
    \left\langle y^2\right\rangle
    =\sigma_y^2=\frac{1+e^{2r}}{2}.
    \end{equation}
    We introduce the normalized sampling variables 
	\begin{equation}\label{Eq:normalized_quadratures}
		\bar{x}
		=
		\frac{x}{\sqrt{2}\cosh r},
		\qquad
		\bar{y}
		=
		\frac{y}{\sqrt{2} \cosh r},
	\end{equation}
	so that
	\begin{equation}
		\left\langle
		\bar{x}^2+\bar{y}^2
		\right\rangle
		=
		1.
	\end{equation}
	Then, the temporal kernel of the $i$-th BSV realization is written as
	\begin{equation}\label{Eq:K_BSV_revised}
		K_{\mathrm{BSV}}^{(i)}(t)
		=
		\bar{x}^{(i)}
		\cos\!\left[
		\omega t+\varphi(\mathbf{s})
		\right]
		+
		\bar{y}^{(i)}
		\sin\!\left[
		\omega t+\varphi(\mathbf{s})
		\right],
	\end{equation}
	and the corresponding spatiotemporal BSV field is
	\begin{equation}\label{Eq:BSV_Est_revised}
		E_{\mathrm{BSV}}^{(i)}(\mathbf{s},t)
		=
		\left|\widetilde{u}(\mathbf{s})\right|
		K_{\mathrm{BSV}}^{(i)}(t)
		f(t).
	\end{equation}
	Here, $\left|\widetilde{u}(\mathbf{s})\right|$ is given by the modulus of the spatial field in Eq.~\eqref{Eq:Gaussian_mode_revised}, and the mean peak intensity is given by $\langle I_0 \rangle=\varepsilon_0cE^2_0/2$.
	For the hybrid two-color driving field, the electric field of the $i$-th
	realization is expressed as
	\begin{equation}\label{Eq:two_color_field_revised}
		\begin{aligned}
			E_{\mathrm{TC}}^{(i)}(\mathbf{s},t)
			&=
			\left|\widetilde{u}_1(\mathbf{s})\right|
			\cos\!\left[
			\omega_1t
			+\varphi_1(\mathbf{s})
			\right]f(t)
			\\
			&+
			\left|\widetilde{u}_2(\mathbf{s})\right|
			\left\{
			\bar{x}^{(i)}
			\cos\!\left[
			\omega_2 t+\varphi_2(\mathbf{s})
			\right]
			+
			\bar{y}^{(i)}
			\sin\!\left[
			\omega_2 t+\varphi_2(\mathbf{s})
			\right]
			\right\}
			f(t),
		\end{aligned}
	\end{equation}
	where the subscripts 1 and 2 denote the coherent and BSV fields, respectively. $f(t)$ is temporal envelope of the coherent and BSV pulses, respectively.

    It is emphasized that semiclassical sampling approximation requires particular care at small squeezing parameters. For a single-mode squeezed vacuum, the mean photon occupation decreases with decreasing $r$, and the vacuum contribution inherent in the Husimi-$Q$ distribution becomes comparable to or larger than the photon-number contribution. Treating the sampled amplitudes as classical driving fields can therefore introduce substantial deviations from the quantum dynamics. 
    For example, for the vacuum state $r = 0$, the samples could still produce strong HHG emission while it contradicts the physical reality.
    Thus, the present semiclassical sampling procedure is intended for the bright-field regime, i.e. large squeezing parameter values.

	\subsection{Calculation of high-order harmonic spectra}\label{sec:2_2}
	
	Although the BSV field exhibits quantum fluctuations distinct from those of a coherent field, each individual field realization has a well-defined waveform\cite{Kern2026}. This allows each sampled field realization to be treated using well-established theories of coherent-field-driven HHG to investigate the yield at both the microscopic and macroscopic levels.
	The central idea is to represent the quantum-field fluctuations by an ensemble of coherent-field realizations. For each sampled driving field $E^{(i)}_{\mathrm{TC}}$, the corresponding complex harmonic field $\widetilde{E}^{(i)}_{\mathrm{h}}$ is calculated independently. Finally, the mean HHG spectrum is then obtained by the incoherent average of all realizations $\langle |\widetilde{E}^{(i)}_{\mathrm{h}}|^2\rangle$.
	
	Here, we consider HHG driven by a linearly polarized laser pulse. At the microscopic level, the single-atom HHG spectrum $|\widetilde{E}^{(i)}_{\mathrm{h}}|^2$ is obtained by (in atomic units)
	\begin{equation}\label{Eq:hhg_sar}
		|\widetilde{E}^{(i)}_{\mathrm{h}}(\omega)|^2 = \frac{2\omega^4}{3\pi c^3} |\widetilde{D}^{(i)}(\omega)|^2,
	\end{equation}
	where $\widetilde{D}^{(i)}(\omega)$ is the Fourier transform of the induced dipole $D^{(i)}(t)$ in length gauge. The dipole $D^{(i)}(t) = \langle \Psi^{(i)}|\textbf{r}| \Psi^{(i)}\rangle$ is calculated by solving the one-dimensional time-dependent Schr\"odinger equation (1D-TDSE) for each sample
	\begin{equation}
		i\frac{\partial}{\partial t}\Psi^{(i)}(x,t) = \hat{H}^{(i)}(t)\Psi^{(i)}(x,t),
	\end{equation}
	with the Hamiltonian given in atomic units as
	\begin{equation}
		\hat{H}^{(i)}(t) = -\frac{1}{2}\frac{\partial^2}{\partial x^2} + V_{\mathrm{atom}}(x) + V_{\mathrm{abs}}(x) + xE^{(i)}_{\mathrm{TC}}(t).
	\end{equation}
	In this work, we use krypton (Kr), for which the atomic potential is represented by a short-range P\"oschl--Teller potential
	\begin{equation}
		V_{\mathrm{atom}}(x) = -\frac{V_0}{\cosh^2(\alpha x)},
	\end{equation}\(V_{\mathrm{abs}} = -i\eta(|x|-R_{\mathrm{abs}})^3\Theta(|x|-R_{\mathrm{abs}})\) is a cubic absorbing potential to prevent boundary reflections, with grid length \(L\), \(\eta=10^{-4}\) a.u., and \(R_{\mathrm{abs}}=0.4L\).
	In principle, the induced dipole can be calculated using established theoretical approaches, such as the strong-field approximation (SFA)\cite{Le2016}  or quantitative rescattering theory (QRS)\cite{Le2009}. Because the single-atom response must be evaluated for a large number of BSV realizations, direct TDSE calculations would be computationally expensive. We therefore use a GPU-accelerated deep residual multilayer perceptron (ResMLP) as a surrogate model. The model bypasses the computationally expensive wavefunction propagation, directly mapping the physical driving parameters to the time-dependent induced dipole. This enables the rapid evaluation of the microscopic dipole response with high accuracy and speeds up the computation by over three orders of magnitude.

	To calculate macroscopic HHG response, Maxwell's equations coupled to the microscopic nonlinear polarization must be solved to account for propagation effects. The theory of macroscopic HHG driven by coherent fields has been well established and extensively validated in experiments\cite{Sun2017,Fu2022,Li2025b,Li2026a}. However, for the ensemble considered here, the computational cost is prohibitive to perform the full three-dimensional propagation calculations incorporate full propagation effects for every realization in the ensemble. We therefore introduce several approximations to reduce the computational cost while retaining the essential physical mechanism. 
	
	i) the propagation of the harmonic field is described using the following one-dimensional approximation
    	\begin{equation}\label{Eq:mac_HHG_prop}
		-\frac{2i\omega}{c}\frac{\partial \widetilde{E}^{(i)}_{\mathrm{h}}(z,\omega)}{\partial z}
		=\frac{\omega^{2}}{c^{2}}( 1 - \tilde{n}^2)\widetilde{E}^{(i)}_{\mathrm{h}}(z,\omega)
		-\mu_{0}\omega^{2}\widetilde{P}^{(i)}_{\mathrm{NL}}(z,\omega).
	\end{equation}
	In this approximation, radial diffraction is neglected, whereas neutral-gas dispersion and photoabsorption are retained. Here, $\mu_0$ is the vacuum permeability, $c$ is the speed of light, $\tilde{n} = 1+\delta_{\mathrm{h}} -i\beta_{\mathrm{h}}$ is the complex refractive index of the gaseous medium where $\delta_{\mathrm{h}}$ and $\beta_{\mathrm{h}}$ account for dispersion and absorption, respectively. Second, the plasma-induced modification of the driving fields is neglected. In this case, the driving field retain their prescribed profiles within this approximation so that the microscopic response of HHG is accessible in our AI-accelerated method. The nonlinear polarization
	\begin{equation}
		\widetilde{P}^{(i)}_{\mathrm{NL}}(z,\omega)
		=\hat{\mathcal{F}}\left\{ N_{0} D^{(i)}(z,t) \right\}
	\end{equation}
	is calculated from the local driving field $E^{(i)}_{\mathrm{TC}}(z,t)$ of driving laser. $\hat{\mathcal{F}}$ denotes the Fourier transform, and $N_0$ is the atomic number density. 
	
	ii) For the driving fields, we adopt the spatial form $\widetilde{u}^{j}(z)$ as
	\begin{equation}\label{Eq:oneD_mode_driving}
		\widetilde{u}^{j}(z) = E^j_0 \exp\left[i (\varphi^{j}_{\mathrm{geo}} + \varphi^{j}_{\mathrm{dis}})\right] / \sqrt{1+(z/z^{j}_{\mathrm{R}})^2},
	\end{equation}
	based on the analytical Gaussian form Eq.~\eqref{Eq:Gaussian_mode_revised} but only considering the on-axis part ($r=0$). Here, $j$ represents the 800-nm or 1600-nm field. It includes both the Gouy phase shift $\varphi^{j}_{\mathrm{geo}}$ and neutral-gas dispersion $\varphi^{j}_{\mathrm{dis}}$. The corresponding phase contributions are given by
	\begin{equation}\label{Eq:phase_prop}
		\begin{aligned}
			\varphi^{j}_{\mathrm{geo}}(z) &= \arctan (z/z^{j}_{\mathrm{R}}),\\
			\varphi^{j}_{\mathrm{dis}}(z) &= \int^{z}_{z_0} - k^j \left(\delta n^j_1 - 1 \right)dz'.
		\end{aligned}
	\end{equation}
	In Eq.~\eqref{Eq:phase_prop}, $z^{j}_{\mathrm{R}}$ is the Rayleigh length. $\delta n^j_1$ is the refractive index difference between gas medium and vacuum for the driving field. $z_0$ is the entrance position of the gas medium, and $k^j$ is the wave numbers. The phases are calculated for both coherent and BSV fields, and they are included in $\varphi_{j}(\textbf{s})$ in Eq.~\eqref{Eq:two_color_field_revised}.

	\subsection{Phase-matching analysis}\label{sec:2_3}
	
	The phase mismatch $\Delta k$ of the harmonics during macroscopic propagation can be decomposed into four contributions\cite{Weissenbilder2022}
	\begin{equation}
		\Delta k
		=
		\Delta k_{\mathrm{geo}}
		+
		\Delta k_{\mathrm{dis}}
		+
		\Delta k_{\mathrm{ele}}
		+
		\Delta k_{\mathrm{dip}}.
	\end{equation}
	Here, $\Delta k_{\mathrm{geo}}$ is associated with the Gouy phase, while
	$\Delta k_{\mathrm{dis}}$ and $\Delta k_{\mathrm{ele}}$ arise from the neutral-atom and plasma dispersion, respectively. The term $\Delta k_{\mathrm{dip}}$ is primarily determined by the intensity-dependent action phase $\varphi_{\mathrm{act}}$. When the BSV field is sufficiently weak, the first three contributions are approximated by those obtained for the coherent field alone
	\begin{equation}\label{Eq:dk_components}
		\begin{aligned}
			\Delta k_{\mathrm{geo}}(z)
			&\approx
			-\frac{\omega_{\mathrm{h}}}{\omega}
			\frac{1}{z_{\mathrm{R}}}
			\frac{1}{1+(z/z_{\mathrm{R}})^2},
			\\
			\Delta k_{\mathrm{dis}}(z)
			&\approx
			\frac{\omega_{\mathrm{h}}}{c}
			\frac{\rho(\alpha_1-\alpha_{\mathrm{q}})}
			{2\varepsilon_0}
			\left[1-\eta(z)\right],
			\\
			\Delta k_{\mathrm{ele}}(z)
			&\approx
			-\frac{\omega_{\mathrm{h}}}{c}
			\frac{e^2\eta(z)\rho}
			{2\varepsilon_0m_{\mathrm{e}}\omega^2},
		\end{aligned}
	\end{equation}
	where $\hbar\omega_{\mathrm{h}}$ is the harmonic photon energy, while $\omega$ and $z_{\mathrm{R}}$ are the angular frequency and Rayleigh length of the coherent field, respectively. The quantities $\alpha_1$ and $\alpha_{\mathrm{q}}$ are the atomic polarizability at the fundamental and harmonic frequencies, respectively, and $\eta$ is the ionization fraction of the medium. $e$ is the electron charge, and $m_{\mathrm{e}}$ is the electron mass. $\varepsilon_0$ is the vacuum permittivity. The neutral-gas and plasma contributions, $\Delta k_{\mathrm{dis}}$ and $\Delta k_{\mathrm{ele}}$, are proportional to the gas pressure $P$ through
	\begin{equation}
		\rho=\frac{P}{k_{\mathrm{B}}T},
	\end{equation}
	where $\rho$ is the atomic number density, $k_{\mathrm{B}}$ is the Boltzmann constant, and $T$ is the gas temperature.
		The dipole contribution can be approximated as
	\begin{equation}
		\Delta k_{\mathrm{dip}}
		\approx
		\nabla\varphi_{\mathrm{act}},
	\end{equation}
	where the action phase $\varphi_{\mathrm{act}}$ is the time integral of the semiclassical Hamiltonian\cite{Lewenstein1995}. Within the semiclassical strong-field approximation and in atomic units
	$(e=\hbar=m_{\mathrm{e}}=1)$, it is expressed as
	\begin{equation}\label{Eq:action_phase}
		\varphi_{\mathrm{act}}
		=
		\int_{t_{\mathrm{b}}}^{t_{\mathrm{r}}}
		\left[
		\frac{\left[p+A(t)\right]^2}{2}
		+
		I_{\mathrm{p}}
		\right]dt.
	\end{equation}
	Here, $p$ is the canonical momentum of the electron, $A(t)$ is the vector potential of the driving field, and $I_{\mathrm{p}}$ is the ionization potential of the target atom. The ionization and recombination times, $t_{\mathrm{b}}$ and $t_{\mathrm{r}}$, respectively, are determined by solving the classical equations of motion for the electron in the driving field.	
	Treating the weak field as a perturbation, the action phase can be approximated as
	\begin{equation}
		\varphi_{\mathrm{act}}
		\approx
		\varphi_{\mathrm{coh}}+\sigma,
	\end{equation}
	in which
	\begin{equation}\label{Eq:action_phase_decomposition}
		\begin{aligned}
			\varphi_{\mathrm{coh}}
			&=
			\int_{t_{\mathrm{b}}}^{t_{\mathrm{r}}}
			\left[
			\frac{\left[p+A_1(t)\right]^2}{2}
			+
			I_{\mathrm{p}}
			\right]dt,
			\\
			\sigma
			&=
			\int_{t_{\mathrm{b}}}^{t_{\mathrm{r}}}
			A_2(t)
			\left[p+A_1(t)\right]dt.
		\end{aligned}
	\end{equation}
    Here, $\varphi_{\mathrm{coh}}$ represents the action phase accumulated in the coherent field alone, while $\sigma$ is the additional perturbative phase induced by the weak field.
	In deriving Eq.~\eqref{Eq:action_phase_decomposition}, terms of second order in the weak vector potential $A_2(t)$ have been neglected. The vector potentials of strong coherent field and weak perturbing field are $A_1(t)$ and $A_2(t)$, written as
	\begin{equation} \label{Eq:vecpot}
		A_1(t)=-A_0\sin(\omega t),
		\quad
		A_2(t)=-A_{\mathrm{p}}
		\sin\!\left(\frac{\omega t}{2}+\phi_{\mathrm{TC}} \right),
	\end{equation}
	respectively. Note that $\phi_{\mathrm{TC}}=\varphi_2-\varphi_1/2$ is evaluated by 1600-nm field.

    Denoting the pressure-dependent phase-mismatching contributions as
    \begin{equation}
	\Delta k_{\mathrm{dis}}
	+
	\Delta k_{\mathrm{ele}}
	=
	aP,
    \end{equation}
    where $P$ is the gas pressure, and $a$ is given by
    \begin{equation}\label{Eq:pressure_coefficient}
	a
	=
	\frac{\omega_{\mathrm{h}}}
	{2\varepsilon_0 c k_{\mathrm{B}}T}
	\left[
	(1-\eta)
	(\alpha_1-\alpha_{\mathrm{q}})
	-
	\frac{\eta e^2}
	{m_{\mathrm{e}}\omega_1^2}
	\right].
    \end{equation}
    Equation~(\ref{Eq:pressure_coefficient}) is obtained by substituting
    $\rho=P/(k_{\mathrm{B}}T)$ into the neutral-gas and plasma contributions in Eq.~\eqref{Eq:dk_components}. We write the phase-matching gas pressure in the presence of the weak field as
    \begin{equation}
	P_{\mathrm{PM}}
	=
	P_0+\Delta P,
    \end{equation}
    where $P_0$ is the phase-matching gas pressure in the strong coherent field alone
    \begin{equation}
 	\left(
	\Delta k_{\mathrm{geo}}
	+
	aP_0
	+
	\nabla\varphi_{\mathrm{coh}}
	\right)
	=0,   
    \end{equation}
    and $\Delta P$ is the pressure shift induced by the weak field. The phase-matching condition can then be expressed as
    \begin{equation}\label{Eq:pressure_phase_matching}
	\left(
	\Delta k_{\mathrm{geo}}
	+
	aP_0
	+
	\nabla\varphi_{\mathrm{coh}}
	\right)
	+
	\left(
	a\Delta P
	+
	\nabla\sigma
	\right)
	=
	0.
    \end{equation}
    Thus, when the weak 1600-nm field is present, an additional pressure shift
    \begin{equation}\label{Eq:p_offset}
	\Delta P
	=
	-\frac{\nabla\sigma}{a}
    \end{equation}
    is required to compensate for the additional phase mismatch induced by $\sigma$. Substituting Eq.~\eqref{Eq:vecpot} into the expression for $\sigma$ in Eq.~\eqref{Eq:action_phase_decomposition} gives
    \begin{equation}\label{Eq:sigma_explicit}
	\sigma
	=
	-A_{\mathrm{p}}
	\int_{t_{\mathrm{b}}}^{t_{\mathrm{r}}}
	\left[
	p-A_0\sin(\omega t)
	\right]
	\sin\!\left(
	\frac{\omega t}{2}+\phi_{\mathrm{TC}}
	\right)dt.
    \end{equation}
    Here, $A_{\mathrm{p}}$ is the amplitude of the 1600-nm vector potential, and $\phi_{\mathrm{TC}}$ is the relative phase between two-color fields. Thus, it is concluded that the phase-matching condition is affected by both the weak-field intensity and the relative phase.

	\section{Results and Discussion}

    \subsection{Pressure-dependent harmonic yields of spectral components}\label{sec:3_1}
    
\begin{figure}[b!]
  	\centering
  	\includegraphics[width=0.95\linewidth]{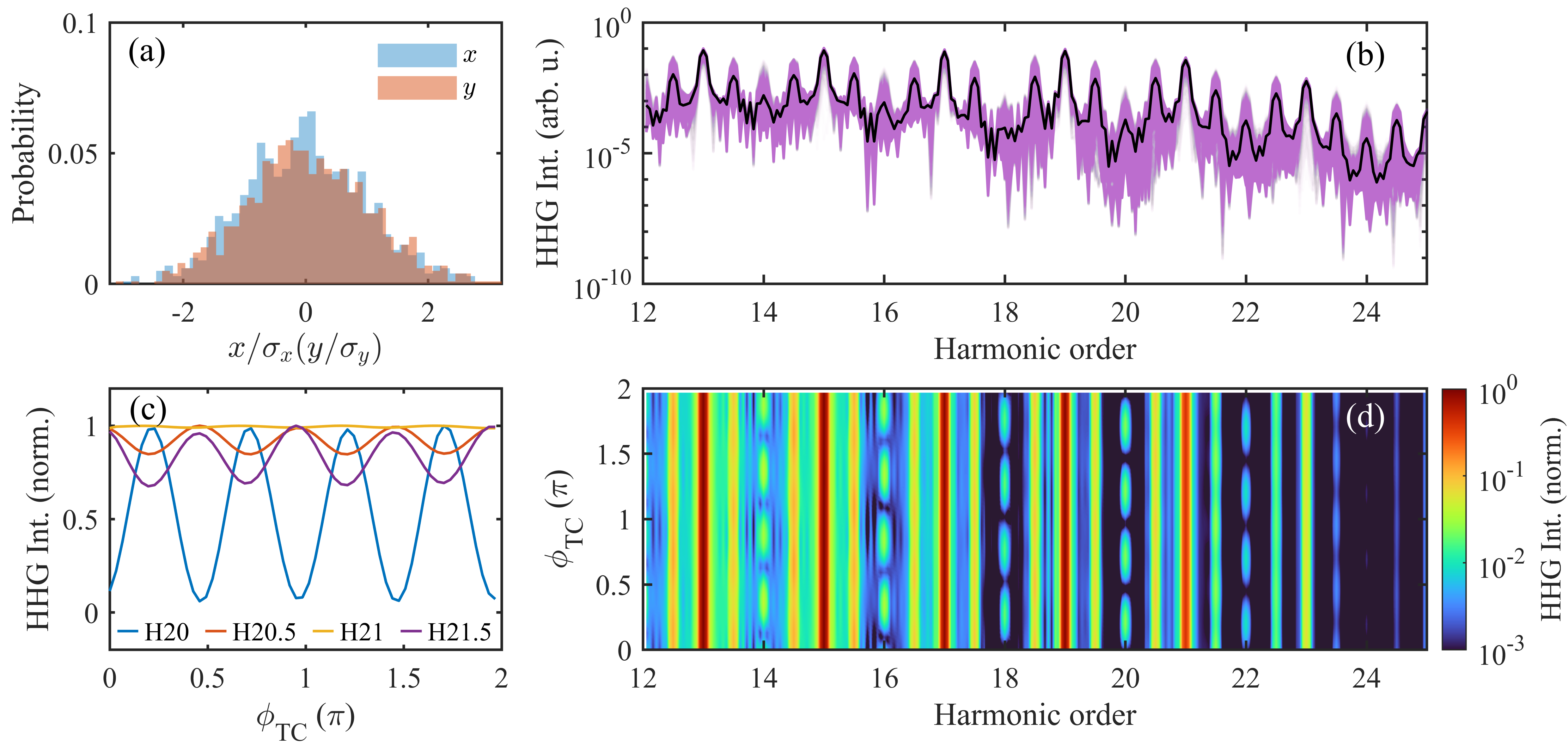}
  	\caption{Single-atom high-order harmonic spectra driven by hybrid two-color light quantum field. 
  		(a) Probability distribution of orthogonal components $x$ and $y$ in simulation. 
  		(b) Shot-by-shot high-order harmonic spectra under the relative phase of 0. The black line indicates the averaged (incoherent summation) mean intensity.
  		(c) Harmonic yield of four spectral components as a function of the relative phase $\phi_{\mathrm{TC}}$ (defined in terms of the period of 1600-nm field) between two fields.
  		(d) High-order harmonic spectra as a function of the relative phase between two fields.} 
  	\label{fig:single_atom_HHG}
  	\end{figure}

    We first calculate the HHG spectra at the single-atom level. The peak intensity of 800-nm field is $1.5 \times 10^{14}~\mathrm{W/cm^2}$, and mean intensity of 1600-nm BSV field is $1.5 \times 10^{11}~\mathrm{W/cm^2}$.A trapezoid temporal envelope (3-o.c. rise, 8-o.c. plateau, 3-o.c. fall) is adopted, preceded by a 2-o.c. field-free evolution, totaling an 18-o.c. simulation. The BSV field is simulated using Eq.~\eqref{Eq:K_BSV_revised} with 1000 random samples, whose orthogonal components $x$ and $y$ follow Gaussian distributions as shown in Fig.~\ref{fig:single_atom_HHG}(a). In the calculations, the squeezing parameter $r=15$ is used such that $\sigma_y/\sigma_x >10^6$. Under these conditions, the simulated BSV field exhibits intensity fluctuation and phase ambiguity of $\pi$.
    Fig.~\ref{fig:single_atom_HHG}(b) shows the single-atom HHG spectra when relative phase between two fields is 0. Individual realization of HHG is plotted in purple, and their ensemble-averaged spectrum is shown in black. The harmonic yields exhibit shot-to-shot fluctuations, especially for even and half-integer sidebands components.
    Fig.~\ref{fig:single_atom_HHG}(d) shows the HHG spectra as a function of the relative phase between two fields, defined by the period of 1600-nm field. The harmonic yield varies with a period of $\pi/2$, which has been investigated and explained by a four-slit interference model\cite{Tzur2025a}.
    
    Notably, the harmonic yields of H$_{2N\pm1/2}$ are comparable in the plateau below H20, and generally about an order of magnitude weaker than odd-order ones.
    Fig.~\ref{fig:single_atom_HHG}(c) shows the harmonic yield as a function of relative phase for the four consecutive components H20, H20.5, H21 and H21.5. The modulation trend is similar within integer and half-integer components, i.e. the relative phase when the maximum or minimum occurs is very close.

\begin{figure}[t]
        \centering
        \includegraphics[width=0.95\linewidth]{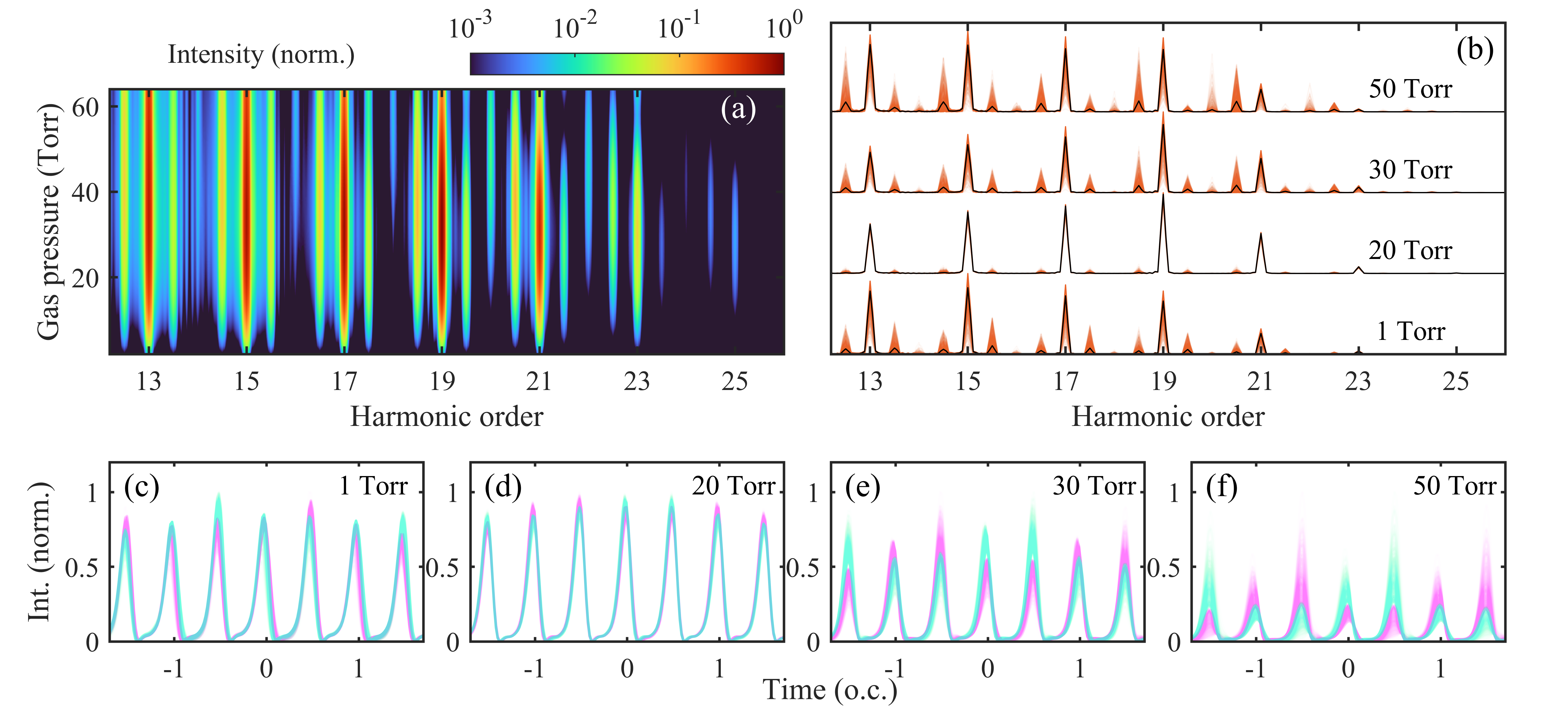}
        \caption{Macroscopic high-order harmonic generation driven by the two-color quantum light field. 
        (a) Mean high-order harmonic intensity at different gas pressures, normalized to the global maximum.
        (b) Shot-by-shot spectra (brown) and ensemble-averaged spectrum (black) at four gas pressures, plotted on a linear scale. 
        (c-f) Shot-by-shot synthesized pulses correspond to the gas pressures in panel (b). For illustration, pulses generated for BSV field with phase of 0 ($\pi$) are indicated by pink (cyan) respectively.}
        \label{fig:mac_QHHG}
    \end{figure}

    Then we perform the macroscopic propagation simulation. The driving-pulse parameters are the same as those used in Fig.~\ref{fig:single_atom_HHG}(b). The beam waists are 60 $\mu$m for the 800-nm field and 25 $\mu$m for the 1600-nm, respectively. The gas medium of krypton is 1.5 mm long, with its entrance placed at 3.5 mm before the laser focus. The mean macroscopic HHG spectra as a function of gas pressure are shown in Fig.~\ref{fig:mac_QHHG}(a). The simulations show that the gas pressures for the maximum harmonic yield of four spectral components are very different. The neutral atom dispersion of 800-nm and 1600-nm field are very close, although it introduces an extra phase of driving field. Thus, it is supposed not to change the relative phase between two fields significantly. In addition, the intensity modulation between H$_{2N\pm1/2}$, H$_{2N}$ and H$_{2N+1}$ are quite different, which is distinguished from that at the single-atom level as shown in Fig.~\ref{fig:single_atom_HHG}(c). The pressure-dependent harmonic intensity can also be interpreted as both the spectral structure and intensity being changed with gas pressure. 
    Specifically, the changes of spectral structure are shown in Fig.~\ref{fig:mac_QHHG}(b). 
    At 1 Torr, the even harmonics are very weak, and H$_{2N\pm 1/2}$ components are close to each other. 
    At 20 Torr, the odd harmonics become dominant so that other components are hardly observed. 
    At 30 Torr, the half-integer sidebands harmonics become significant again. At the same time, even harmonics like H20 and H22 are also observed although their mean intensities remain low. 
    At 50 Torr, H$_{2N+1/2}$ is more intense than H$_{2N-1/2}$ overall. In addition, due to the large intensity fluctuation, it can be even comparable to the main odd harmonic. And the H$_{2N}$ harmonics also become more prominent.

    The corresponding time-domain pulse trains are shown in Fig.~\ref{fig:mac_QHHG}(c)-(f). In this $\omega+\omega/2$ configuration, the $\pi$ phase change of the $\omega/2$ field is equal to a translation by one optical cycle of the fundamental $\omega$ field. For HHG process, the $0-\pi$ phase ambiguity is indistinguishable in terms of the spectral intensity. Thus, for visual guidance, we plot the pulses with two colors to distinguish those generated from positive (purple) or negative (cyan) amplitude $y$ in Eq.~\eqref{Eq:K_BSV_revised}. Then, the temporal profiles provide a complementary representation of the spectral changes. It can be clearly seen that the intensities of sub-cycle bursts in the pulse train differ significantly with gas pressure. Notably, the intensity of sub-cycle bursts differ less from each other at 20 Torr--resembling a single-color case. This corresponds to the spectra dominated by odd harmonics in Fig.~\ref{fig:mac_QHHG}(b) at 20 Torr.

    \subsection{Amplitude asymmetry modulated spectral structure}\label{sec:3_2}
    
\begin{figure}[t!]
    \centering
    \includegraphics[width=0.95\linewidth]{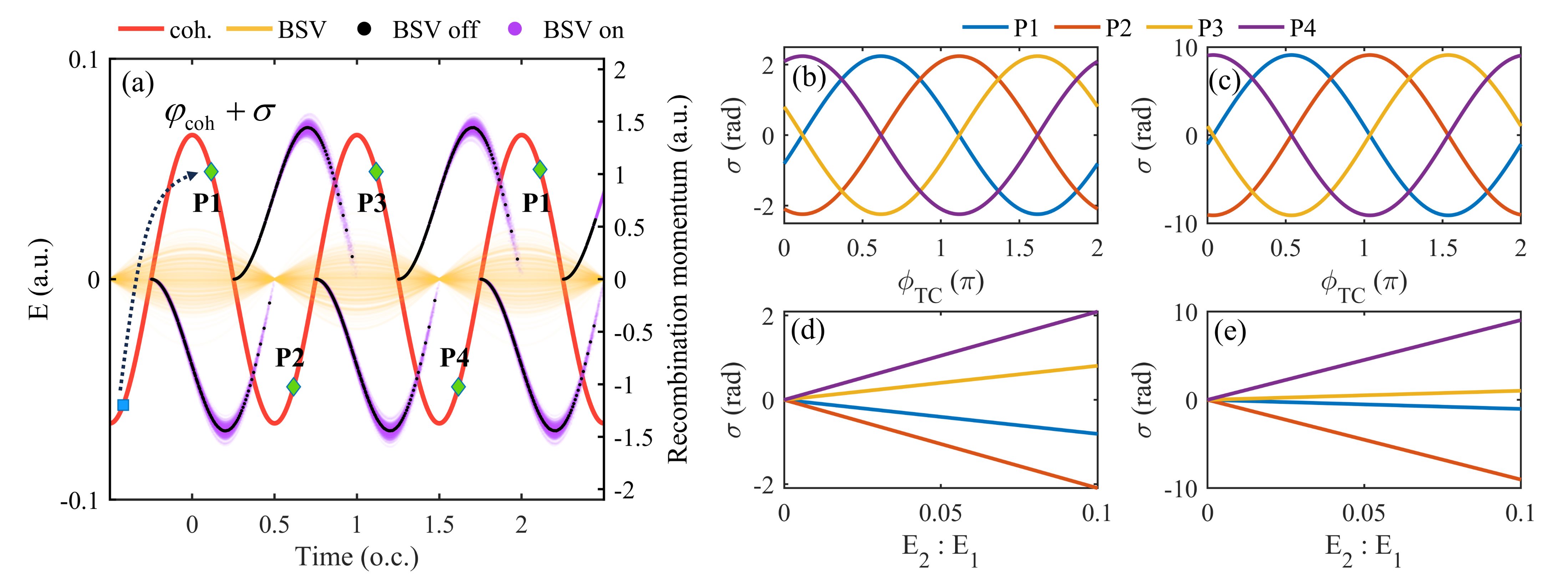}
    \caption{Four-channel interference picture of the harmonic generation. 
    (a) Illustration of the two-color quantum driving field (lines, left axis) and recombination momentum (dots, right axis) of electron by solving the classical motion equation. The BSV field is multiplied by a constant for visual guidance. P1-P4 denote four contributions with the same recombination energy in every two optical cycles (o.c.) of the coherent field. 
    (b,c) The perturbation of action phase $\sigma$ with respect to short- and long-trajectory electron, in addition to that when coherent 800-nm field exists only. The amplitude ratio is $E_2:E_1 = 0.1$ and $\phi_{\mathrm{TC}}$ refers to their relative phase, defined by period of 1600-nm field. 
    (d,e) The same as (b,c) except that $\phi_{\mathrm{TC}}=0$ and amplitude ratio is varied. $\sigma$ presented in (b-e) are illustrated in a coherent realization.
    }\label{fig:four_wave_model}
    \end{figure}

    To understand the mechanism of spectral modulation in macroscopic propagation, the temporal interference picture is exploited. Fig.~\ref{fig:four_wave_model}(a) illustrates the electric field of both coherent and BSV components and the corresponding time-momentum relation of electron recollision. In the presence of the 1600-nm BSV field, due to the period of driving field, there are four distinguishable recollision events responsible for the harmonic emission at a given photon energy (labeled by P1 - P4) in every two optical cycles of the 800-nm field. As shown by the dots, although the weak BSV field perturbs the recombination time and momentum (purple), it would not destroy the typical feature in a single-color coherent field (black). Since the time integral of Hamiltonian, namely action phase, is related to the dipole phase, we approximate the action by $\varphi_{\mathrm{act}} \approx \varphi_{\mathrm{coh}} + \sigma$. $\varphi_{\mathrm{coh}}$ represents the one obtained in the 800-nm field only, and $\sigma$ is the additional term contributed by BSV field (See details in Sec. \ref{sec:2_3}). According to Eq.~\eqref{Eq:sigma_explicit}, $\sigma$ depends on the amplitude of 1600-nm field, and relative phase between two fields. We calculated the $\sigma$ of short- and long-trajectory electrons in a coherent realization as shown in Fig.~\ref{fig:four_wave_model}(b)-(e).
    First, $\sigma$ varies sinusoidally with the relative phase approximately, and sequentially successive phase-perturbation curves are shifted by approximately $\pi/2$ from P1 - P4. This yields the relation $\sigma_1 = -\sigma_3;\sigma_2 = -\sigma_4$.
    Second, for a certain relative phase, $\sigma$ is proportional to the amplitude of the 1600-nm field, and $\sigma$ of long-trajectory electrons is larger than that of short-trajectory ones.
    It is noted that the relation above holds for BSV field although it exhibits shot-to-shot fluctuations.

    The harmonic field can be approximated by
    \begin{equation}
    	\tilde{H}_q\propto\sum_{m=1}^{4} (-1)^{m+1}\cdot A_m\cdot \exp\left\{i\left[(m-1)q\pi+\sigma_m\right]\right\},
    \end{equation}
    containing the four contributing emissions P1 - P4. Here, the global phase factor is omitted. $A_m$ and $\sigma_m$ represent the amplitude and phase perturbation of sub-cycle emission that contributes to the final $q$-th harmonic. Substituting $q=2N\pm1/2
    $, the complex field of half-integer sidebands harmonics $\widetilde{\mathrm{H}}_{2N\pm1/2}$ can be expressed as
    \begin{equation}
    \widetilde{\mathrm{H}}_{2N\pm1/2} \propto (A_1 e^{i\sigma_1}-A_3 e^{i\sigma_3}) \mp i (A_2 e^{i\sigma_2}-A_4 e^{i\sigma_4})
    \end{equation}
    With the relation $\sigma_1 = -\sigma_3$ and $\sigma_2 = -\sigma_4$, the spectral intensity is further written as 
    \begin{equation}
    |\widetilde{\mathrm{H}}_{2N\pm1/2}|^2 \propto A_{\mathrm{BG}} \pm \Delta,
    \end{equation}
    consisting of a common background term
    \begin{equation}
    \begin{aligned}
    A_{\mathrm{BG}} & = (A_1-A_3)^2\cos^2\sigma_1 + (A_1+A_3)^2\sin^2\sigma_1\\ &+ (A_2-A_4)^2\cos^2\sigma_2 + (A_2+A_4)^2\sin^2\sigma_2
    \end{aligned}
    \end{equation}
    and the modulation term
    \begin{equation}\label{eq:spec_modulation_term}
    \Delta = 2(A_1+A_3)(A_2+A_4)\left[\epsilon_{13}\cos\sigma_1\sin\sigma_2 -\epsilon_{24}\cos\sigma_2\sin\sigma_1 \right],
    \end{equation}
    where $\epsilon_{13}=(A_1-A_3)/(A_1+A_3)$ and $\epsilon_{24}=(A_2-A_4)/(A_2+A_4)$ reflect the amplitude asymmetry of the sub-cycle contributions. Thus, there are two factors that determine the modulation term. One is the phase perturbation $\sigma_1$ and $\sigma_2$, and the other is amplitude asymmetry $\epsilon$. At the single-atom level, $\epsilon$ results from the differences in ionization amplitudes between half cycles--which could be very small since the 1600-nm field is weak. In this case, the intensity of H$_{2N\pm1/2}$ components could be close as shown in Fig.~\ref{fig:single_atom_HHG}(b).
    However, macroscopic propagation can enhance the amplitude asymmetry of the sub-cycle contributions, thereby modifying the phase-dependent interference term and the relative sideband yields. For example, the asymmetry is significant at 50 Torr as shown in Fig.~\ref{fig:mac_QHHG}(f), so a large difference of harmonic intensity between H$_{2N\pm1/2}$ occurs. However, at 1 Torr, the amplitude asymmetry $\epsilon$ is smaller such that the modulation term $\Delta$ is not as pronounced. 

    \subsection{BSV-induced fluctuation of phase-matching condition}\label{sec:3_3}
    
\begin{figure}[t!]
    	\centering
    	\includegraphics[width=0.95\linewidth]{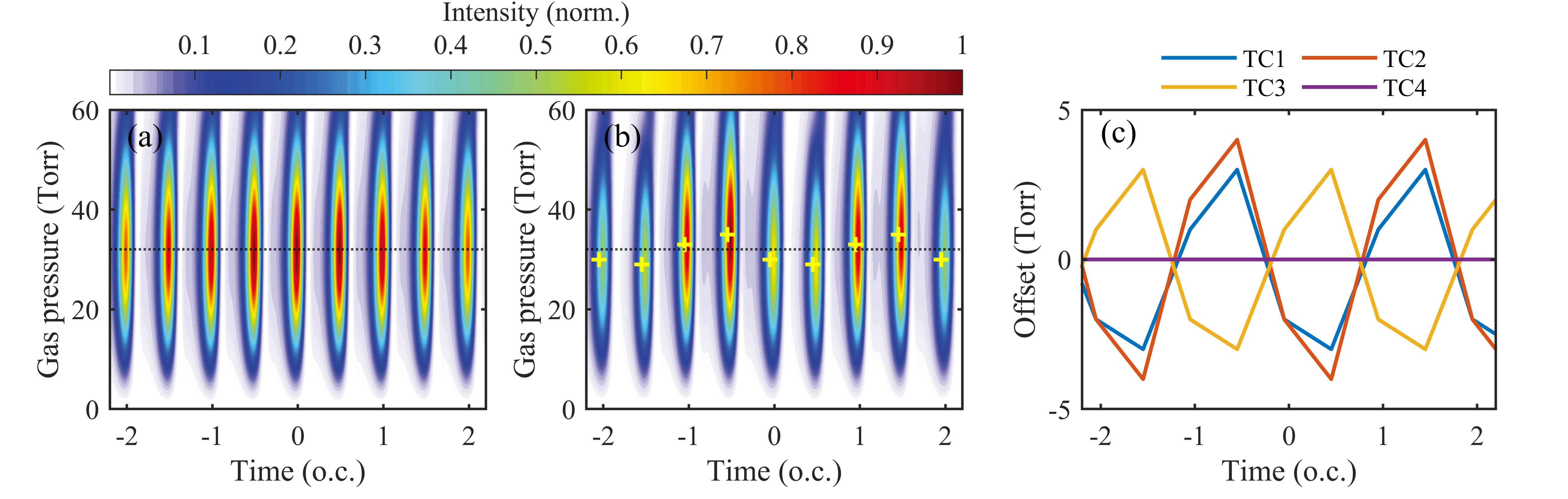}
    	\caption{Subcycle-resolved phase-matching condition.
    		Synthesized pulses by high-order harmonics where 1600-nm is (a) excluded and (b) included. Black dotted line indicates the gas pressure for the optimum yield in sole 800-nm field. Yellow crosses mark the optimal ones for individual sub-cycle bursts. (c) Offset of sub-cycle optimal gas pressures apart from that in sole 800-nm field. Four situations are shown: TC1, same as (b). TC2, enlarge the intensity of 1600-nm field by a factor of two. TC3, change the phase of 1600-nm field by $\pi$. TC4, beam waist of 1600-nm field is twice that of the 800-nm field.}
    	\label{fig:shot_prop_evo}
    \end{figure}

    The pressure-dependent asymmetry implies a time-dependent phase-matching regime. To elucidate this mechanism, we next analyze the phase-matching condition. In Fig.~\ref{fig:shot_prop_evo}(a) and (b), we show the attosecond pulses synthesized by macroscopic HHG at different gas pressures without and with the 1600-nm field, respectively. Both driving fields in this comparison are coherent and peak intensity of 1600-nm field is $1.5 \times10^{11}~\mathrm{W/cm^2}$. In the absence of 1600-nm field, for each sub-cycle burst, the optimal gas pressure for the yield is around 32 Torr, as indicated by the black dotted line. When the 1600-nm field exists, it is observed that the optimal gas pressure (yellow crosses) of each burst has an offset. And the offsets exhibit a sine-like feature along with time and change sign every one optical cycle of 800-nm field, which is similar to that of phase perturbation $\sigma$.
    
    We write the phase-matched gas pressure in the presence of weak 1600-nm field as
    \begin{equation}
	P_{\mathrm{PM}}
	=
	P_0+\Delta P,
    \end{equation}
    $P_0$ is the phase-matching gas pressure in the absence of 1600-nm field, and $\Delta P$ is the offset. Since the $\Delta P$ is proportional to the gradient of $\sigma$ (Eq.~\eqref{Eq:p_offset}), it also depends on the amplitude of 1600-nm field and relative phase between two fields, as well as their gradient (Eq.~\eqref{Eq:sigma_explicit}). In Fig.~\ref{fig:shot_prop_evo}(c) we show the time-dependent pressure offset calculated for the other three cases. Enlarging the intensity of 1600-nm field would increase the offset (TC2). Changing the phase of 1600-nm by $\pi$ flips the offset, while approximately preserving its magnitude (TC3). Notably, when the 1600-nm field is focused much more loosely than the 800-nm field, the pressure offset can be suppressed mostly (TC4). In this case, both $\nabla A_p$ and $\nabla\phi_{\mathrm{TC}}$ are small, so the pressure offset is negligible.
    The optimal gas pressure offset of sub-cycle bursts  will yield different phase mismatch of sub-cycle burst. At a certain gas pressure, the sub-burst coherence length $L_{\mathrm{coh}}=\pi/|\Delta k|$ can be different, which will cause different buildup behaviors of harmonics in the gas medium. Given the same propagation distance, a longer coherence length favors the macroscopic building up of harmonics more. For example, the optimal gas pressure for sub-cycle burst near $t=1.5$ o.c. is highest, i.e. closer to 50 Torr as shown in Fig.~\ref{fig:shot_prop_evo}(c)--meaning that the corresponding burst has the longest coherence length, hence a stronger amplification. This is demonstrated by the pulses shown in Fig.~\ref{fig:mac_QHHG}(f), where the bursts near $t=1.5$ and $t=-0.5$ o.c. are much more dominant than others. 

\begin{figure}
        \centering
        \includegraphics[width=0.95\linewidth]{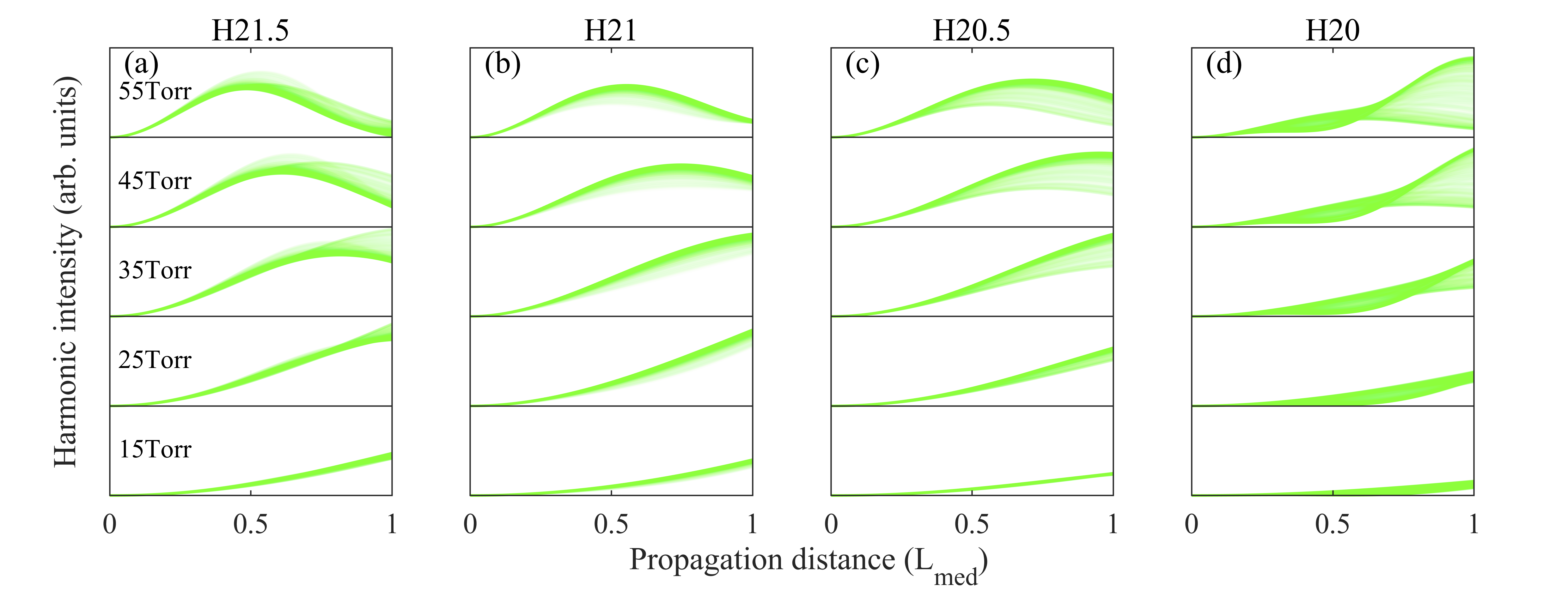}
        \caption{Buildup of high-order harmonics in the gas medium. Four spectral components (a) H21.5, (b) H21, (c) H20.5 and (d) H20 are plotted on a linear scale for different gas pressures. For each row, the gas pressures are the same and labeled in insets of (a). For the shots in each spectral component, they are normalized to their maximum individually to illustrate the phase-matching condition. }
        \label{fig:buildup}
    \end{figure}

    Since the spectral structure is related to the pulse asymmetry and each sub-cycle burst has its own phase-matching condition, the harmonics from different spectral components would also exhibit distinct buildup behaviors. In Fig.~\ref{fig:buildup}, we show the harmonic intensity as a function of propagation distance at different gas pressures for the four spectral components. First, it is observed the buildup dynamics vary between realizations due to the quantum-fluctuation perturbed phase-matching condition as discussed above. For H21.5 at 15 Torr, the buildup curves exhibit a smaller spread while the phase-matching condition is not optimal so that the harmonic intensity doesn't reach the optimum. At 25 Torr, most of the shots are close to a high yield, and the mean intensity achieves its maximum as shown in Fig.~\ref{fig:mac_QHHG}(a). At 35 Torr, many curves begin to decrease before the medium exit, whereas others continue to grow. As the gas pressure increases further, it becomes difficult to maintain favorable phase matching for all shots throughout the medium. Thus, the mean harmonic yield decreases rapidly. The same pattern with growing gas pressure can be also observed for H21 and H20.5. At the gas pressure for optimal harmonic yields, near 32 Torr and 38 Torr respectively, the phase-matching condition among shots are mostly converged to the optimum. While for H20, the shot-to-shot behaviors are more different, meaning that probability of the achievement of favorable phase-matching condition is less than others, hence the lower maximum intensity as shown in Fig.~\ref{fig:mac_QHHG}(a). In addition, it is supposed that the fluctuation feature such as photon numbers could be significant modulated via macroscopic propagation. For example, the shot-to-shot intensity of H20 becomes more divergent with increasing in gas pressure, indicating a growing deviation. The change of photon-number statistic with gas pressure has been demonstrated in experiment\cite{Wang2026}.

	\section{Summary}
    In summary, we investigate the phase matching of high-order harmonic generation driven by a two-color quantum light field consisting of an intense coherent field and a weak bright squeezed vacuum at its half frequency. We show the optimal gas pressures of harmonic yield for odd, even, and half-integer harmonics can be much different. The distinct dependence of harmonic yield on gas pressure is interpreted as the amplitude asymmetry of sub-cycle bursts. Such an asymmetry can be more pronounced than that at the single-atom level, indicating the effect of macroscopic propagation. The phase-matching model reveals that perturbation of the weak half-frequency field on the action phase would introduce an extra phase mismatch to sub-cycle harmonic emissions, which is responsible for the amplitude asymmetry of macroscopic high-order harmonics. In addition, the quantum fluctuation of the BSV field also causes shot different phase-matching condition. The gas pressure for the optimum harmonic yield occurs when the favorable phase-matching condition is achieved for those shots with a stronger microscopic response. In addition, since the shot-to-shot buildup can differ from each other significantly, the statistical variable such photon number could show a macroscopic dependence.

	\funding{This work was supported by National Natural Science Foundation of China (grants no. 12450402, 12604518).}	
	
	\data{The data support the findings of this study are available upon reasonable request from the authors.}
    
	\clearpage	
    \bibliographystyle{sciencemag}
	\bibliography{03_NJP_References}	
	
\end{document}